\documentclass[prb,preprint]{revtex4-1}

\usepackage{amsmath}  % needed for \tfrac, \bmatrix, etc.
\usepackage{amsfonts} % needed for bold Greek, Fraktur, and blackboard bold
\usepackage{graphicx} % needed for figures

\renewcommand{\vec}[1]{\mbox{\boldmath $#1$}}

\newcommand*\Del{\mathrm{\Delta}}                 % Roman Delta

\newcommand{\grad}{ {\bf \nabla } }
\newcommand{\m}{ {\ \mathrm m} }
\newcommand{\s}{ {\ \mathrm s} }

\newcommand{\kg}{ {\ \mathrm {kg}} }
\newcounter{saveeqn}

\begin{document}

\title{On the gravitational redshift}

\author{Klaus Wilhelm}
\affiliation{Max-Planck-Institut f\"ur Son\-nen\-sy\-stem\-for\-schung
(MPS), 37077 G\"ottingen, Germany \\
wilhelm@mps.mpg.de}

\author{Bhola N. Dwivedi}
\affiliation{Department of Physics, Indian Institute of Technology
(Banaras Hindu University), Varanasi-221005, India \\
bholadwivedi@gmail.com}

\date{\today}

%%%%%%%%%%%%%%%%%%%%%%%%%%%%%%%%%%%%%%%%%%%%%%%%%%%%%%%%%%%%%%%%%%%%%%%%%%%%%%
\begin{abstract}
The study of the gravitational redshift\,---\,a relative wavelength
increase of $\approx 2 \times 10^{-6}$ was predicted for solar radiation by
Einstein in 1908\,---\,is still an important subject in modern
physics. In a dispute whether or not atom interferometry experiments
can be employed for gravitational redshift measurements, two research teams
have recently disagreed on the physical cause of the shift. Regardless of any
discussion on the interferometer aspect\,---\,we find that both groups of
authors miss the important point that the ratio of gravitational to the
electrostatic forces is generally very small. For instance, the gravitational
force acting on an electron in a hydrogen atom situated in the
Sun's photosphere to the electrostatic force between the proton and the
electron is approximately $3 \times 10^{-21}$.
A comparison of this ratio with the predicted and
observed solar redshift indicates a discrepancy of many orders of magnitude.
Here we show, with Einstein's early assumption of the frequency of spectral
lines depending only on the generating ion itself as starting point, that
a solution can be formulated based on a two-step process in analogy with
Fermi's treatment of the Doppler effect. It provides a sequence
of physical processes in line with the conservation of energy and momentum
resulting in the observed shift and does not employ a geometric description.
The gravitational field affects the release of the photon and not the atomic
transition. The control parameter is the speed of light.
The atomic emission is then contrasted with the gravitational
redshift of matter-antimatter annihilation events.
\end{abstract}

%%%%%%%%%%%%%%%%%%%%%%%%%%%%%%%%%%%%%%%%%%%%%%%%%%%%%%%%%%%%%%%%%%%%%%%%%%%%%%
\maketitle % title page is now complete

%% main text
\section{Introduction} %%%%%%%%%%%%%%%%%%%%%%%%%%%%%%%%%%%%%%%%%%%%%%%%%%%%%%%%%%%%%%%
\label{s.intro}
%% Sect. 1

The study of the gravitational redshift, a relative wavelength increase of
$\Del \lambda/\lambda \approx 2 \times 10^{-6}$ was predicted for solar
radiation by \citet{Ein08} in 1908, is still an important subject in modern
physics \cite{Kol04,Lae09,Choetal,Tur13}.
\citet{Jew96} had found in electric arc spectra:
\begin{quote}
``[...] that the metallic lines were almost invariably displaced toward the
violet, when compared with the corresponding solar lines.''
\end{quote}
At that time---in 1896---a high pressure in the solar atmosphere was erroneously
considered as causing the shift \citep[cf.][]{LoP91}. Measurements of the
gravitational redshift of solar spectral lines are inherently difficult,
because high speeds of the emitting plasmas in the atmosphere of the Sun
lead to line shifts due to the classical Doppler effect. Improved
observational techniques \citep[cf., e.g.][]{LoPetal,Cacetal,TakUen},
have nevertheless established a shift of
%
%% Eq. 1
\begin{equation}
c_0\,\frac{\Del \lambda}{\lambda} \approx 600~\m\,\s^{-1} ~,
\label{shift_600}
\end{equation}
where $c_0 = 299\,792\,458~\m\,\s^{-1}$ is the speed of light in the vacuum
\cite{BIPM} remote from any masses.
This shift is consistent with Einstein's General Theory of Relativity (GTR)
\cite{Ein16}. Together with various other aspects
of GTR---from the deflection of light by a gravitational centre
\cite{Ein11,Ein16,Dys20,Mik59,Shaetal} to Mercury's
perihelion precession \cite{Ver59,Ein15,NobWil,Wil06}, the current attempts
to measure the Lense-Thirring effect \cite{LenThi} on the
planets' motions caused by the solar rotation \cite{Ior05,Ior12}, and the
Shapiro delay \cite{Sha64,Sha71,Kraetal}---the gravitational redshift is one of
the experimental tests of GTR \cite{Wil06}.

Atom interferometry experiments can be used to measure the acceleration of
free fall, see, for instance, \citet{Pet99,Mue10a}.
The same research team has in the meantime
argued that atom interferometry can also perform gravitational redshift
measurements at the Compton frequency.
This claim was criticized as incorrect by \citet{Wol10}
leading to a response in support of the original result~\cite{Mue10b}.
This controversy has continued until recently
\cite{Wol11,Hoh11,Wol12,Hoh12,HohMue}.

\section{Is there a physical process causing the redshift?}%%%%%%%%%%%%%%%%%%
\label{s.cause}
%% Sect. II

One aspect of the dispute between \citet{Mue10a} and \citet{Wol10}
is particularly disturbing and will be analysed here
in some detail: Even after the prediction of the gravitational redshift
by \citet{Ein08} for over a century and the many observational confirmations
mentioned in Section~\ref{s.intro}, there appears to be no
consensus on the physical process(es) causing the shift. This can be
exemplified by two conflicting statements. The first made by \citet{Wol10}
reads:
\begin{quote}
``The situation is completely different for instruments used for testing
the universality of clock rates (UCR). An atomic clock delivers a periodic
electromagnetic signal the frequency of which is actively controlled to
remain tuned to an atomic transition. The clock frequency is sensitive to the
gravitational potential~$U$ and not to the local gravity field
$\vec{g} = \nabla U$. UCR tests are then performed by comparing clocks through
the exchange of electromagnetic signals; if the clocks are at different
gravitational potentials, this contributes to the relative frequency
difference by $\Del \nu/\nu = \Del U/c^2$.''
\end{quote}
Whereas in the second statement it is claimed by \citet{Mue10b}:\footnote{In
this quotation the expression~$h^2$ indicates the square of the clock
separation~\vec{h} and is not related to Planck's constant~$h$ used below.}
\begin{quote}
``We first note that no experiment is sensitive to the
absolute potential~$U$. When two similar clocks at rest in the laboratory
frame are compared in a classical red-shift test, their frequency difference
$\Del \nu/\nu = \Del U/c^2$ is given by
$\Del U = \vec{g}\,\vec{h} + {\cal O}(h^2)$, where $\vec{g} = \grad{U}$
is the gravitational acceleration in the laboratory frame, \vec{h} is the
clock's separation, $c$ is the velocity of light, and ${\cal O}(h^2)$
indicates terms of order $h^2$ and higher. Therefore,
classical red-shift tests are sensitive to \vec{g}, not to the absolute value
of $U$, just like interferometry red-shift tests.''
\end{quote}

The potential at a distance~$r$ from a gravitational centre with
mass~$M$ is constraint in the weak-field approximation for non-relativistic
cases~\cite{LanLif} by
%
%% Eq. 2
\begin{equation}
- c^2_0 \ll U = - \frac{G_{\rm N}\,M}{r} \le 0 ~,
\label{potential}
\end{equation}
where $G_{\rm N}$ is Newton's constant of gravity. The authors of
Ref.~\cite{Wol10} could refer to many publications in their support
\cite{Ein08,Lau20,Sch60,Wil74,Okuetal,SinSam}.
However, it would be required to define explicitly a reference
potential $U_0$. A definition in line with Eq.~(\ref{potential}) would
give $U_0 = 0$ for $r = \infty$.
Experiments on Earth~\cite{PouReb,Craetal,Hayetal,KraLue,PouSni},
in space~\cite{BauWey} and in the Sun-Earth
system~\cite{StJ28,BlaRod,Bra63,Sni72,LoP91,Cacetal,TakUen} have
quantitatively confirmed in this approximation a relative frequency shift of
%
%% Eq. 3
\begin{equation}
\frac{\nu' - \nu_0}{\nu_0} = \frac{\Del \nu}{\nu_0}
\approx \frac{\Del U}{c^2_0} = \frac{U - U_0}{c^2_0}~,
\label{shift}
\end{equation}
where $\nu_0$ is the frequency of a certain transition at~$U_0$ and $\nu'$
the observed frequency there, if the emission caused by the same transition
had occurred at a potential~$U$. The question whether the shift
happens during the emission process or is a result of a
propagation effect is left open by Dicke in the final section
of Ref.~\cite{Dic60}:
\begin{quote}
``To return briefly to the question of the gravitational red shift, it is
concluded that there could be two different red-shift effects. One would be
interpreted in the usual way as a light propagation effect. The other, if it
exists, would be interpreted as resulting from an intrinsic change in an atom
with gravitational potential. The experiment employing an atomic clock in
space would be one way of observing this effect directly, if it exists.''
\end{quote}
There appears to be agreement, however, that the energy of a photon,
$E_\nu = h\,\nu$, with Planck's constant~$h$, does not vary during the
propagation in a static gravitational field---excluding a variation of
$\nu$ with changing~$U$, if $\nu$ is measured against the coordinate or world
time~\cite{Oku00,Okuetal}. This is consistent with the time
dilation of atomic clocks derived from the GTR~\cite{Ein16} and, consequently,
the matter would be settled, if geometric
effects were considered to be an adequate cause of the gravitational
redshift. \citet{Str00} discussed the modification of the
electric potential by gravity in this context.

\citet{Wol10} and \citet{Mue10b} have tried, however,
to explore physical processes that cause the shift; yet both attempts
%do not meet the physical reality,
are problematic
in view of the fact that the gravitational force acting on the
electron in transition is extremely small relative to the internal forces.
This can easily be verified by a comparison of the weak solar gravitational
force~$\vec{K}^\odot_{\rm G}$
acting on the electron in a hydrogen atom in the photosphere of the Sun
with the electrostatic force~$\vec{K}_{\rm E}$:
%
%% Eq. 4
\begin{equation}
\frac{||\vec{K}^\odot_{\rm G}||}{||\vec{K}_{\rm E}||} =
\frac{G_{\rm N}\,M_\odot\,m_{\rm e}}{R^2_\odot}
\left(\frac{e^2}{4\,\pi\,\varepsilon_0\,a_0^2}\right)^{-1} =
\frac{r^\odot_{\rm S}}{2\,R^2_\odot}\,m_{\rm e}\,{c^2_0}
\left(\frac{e^2}{4\,\pi\,\varepsilon_0\,a_0^2}\right)^{-1} =
3.031 \times 10^{-21}
\label{ratio_Sun}
\end{equation}
with
$G_{\rm N} = 6.674 \times 10^{-11}~\m^3\,\kg^{-1}\,\s^{-2}$~;
$M_\odot = 1.989 \times 10^{30}$~kg, the mass and
$R_\odot = 6.960 \times 10^8$~m, the radius of the Sun;
$m_{\rm e} = 9.109 \times 10^{-31}$~kg, the mass of an electron;
$e = 1.602 \times 10^{-19}$~C, the elementary charge;
$\varepsilon_0 = 8.854 \times 10^{-12}~{\rm F}\,\m^{-2}$, the permittivity of
the vacuum; %electric constant
$a_0 = 5.292 \times 10^{-11}$~m, the Bohr radius;
and
$r^\odot_{\rm S} = 2\,G_{\rm N}\,M_\odot/c^2_0 = 2950$~m,
the Schwarzschild radius of the Sun.

The early attempts to measure the gravitational redshift of solar spectral
lines as well as those of the white dwarf star Sirius~B have been reviewed by
\citet{Het80}. In particular, the wrong value of $21~{\rm km}\,\s^{-1}$
published by \citet{Ada25} has been contrasted with the result of
$(89 \pm 16)~{\rm km}\,\s^{-1}$ obtained by \citet{Greetal} for the companion of
Sirius with $R/R_\odot = 0.0078 \pm 0.0002$ and $M/M_\odot = 1.20 \pm 0.25$.
These radius and mass data inserted into Eq.~(\ref{ratio_Sun}) instead of the
solar values give $5.9 \times 10^{-17}$. Mean gravitational redshifts of
$(53 \pm 6)~{\rm km}\,\s^{-1}$ for six white dwarfs in the Hyades
have been measured by \citet{GreTri}.

Even for the very strong gravitational field of the neutron star EXO 0748-676,
for which \citet{Cotetal} found a redshift of
$z = 0.35$ in Fe\,{\sc xxvi} and Fe\,{\sc xxv} as well as in O\,{\sc viii}
lines,
% ---consistent with a mass range of
% $M/M_\odot \approx 1.4~{\rm to}~1.8$ and
% a radius range of $R \approx (9~{\rm to}~12)$~km---
a calculation similar to Eq.~(\ref{ratio_Sun}) yields
%
%% Eq. 5
\begin{equation}
\frac{||\vec{K}^{\rm NS}_{\rm G}||}{||\vec{K}_{\rm E}||} =
\frac{r_{\rm S}}{2\,R^2}\,m_{\rm e}\,c^2_0
\left(\frac{e^2}{4\,\pi\,\varepsilon_0\,a_0^2}\right)^{-1} \approx
2.5 \times 10^{-11}
\label{ratio_NS}
\end{equation}
with $R = 9.15$~km and $r_{\rm S} = 4130$~m for $M = 1.4~M_\odot$. These values
of $R$ and $r_{\rm S}$ lead to the observed redshift of
%
%% Eq. 6
\begin{equation}
z = \left(1 - \frac{r_{\rm S}}{R}\right)^{-\frac{1}{2}} - 1 = 0.35 ~.
\label{Red_shift}
\end{equation}

\citet{Levetal} and \citet{Bow77} discuss whether a spectral feature at
$\approx 400$~keV observed in the Crab Nebula might be the gravitationally
redshifted 511~keV electron-positron annihilation line from the surface of
the pulsar. \citet{Ram84} concluded that the relatively
narrow widths of annihilation lines from gamma-ray bursts indicates emitting
material close to the surface of a neutron star.

A gravitational redshift from galaxies in clusters has also been reported
\cite{Woj11}.

\section{Towards a solution}%%%%%%%%%%%%%%%%%%%%%%%%%%%%%%%%%%%%%%%%%%%%%%%%%
\label{s.solution}
\subsection{Emission of spectral lines}%%%%%%%%%%%%%%%%%%%%%%%%%%%%%%%%%%%%%%
\label{solution_lines}
%% Subsect. III A
The ratios obtained in Eqs.~(\ref{ratio_Sun}) and (\ref{ratio_NS}) support
Einstein's early assumption \cite{Ein08}:
\begin{quote}
,,Da der einer Spektrallinie entsprechende Schwingungsvorgang wohl als
ein intraatomischer Vorgang zu betrachten ist, dessen Frequenz durch das Ion
allein bestimmt ist, so k\"onnen wir ein solches Ion als eine Uhr von
be\-stimm\-ter Frequenzzahl~$\nu_0$ ansehen.''
\end{quote}
(Since the oscillation process corresponding to a spectral line can probably
be seen as an intra-atomic process, the frequency of which is determined
by the ion alone, we can consider such an ion as a clock with a certain
frequency~~$\nu_0$.)

We feel that this view merits to be fully appraised, although \citet{Ein16}
later concluded that (atomic) ``clocks'' would slow down near gravitational
centres.

\citet{Ein17} also emphasized the importance of the momentum transfer during
the absorption or emission of radiation:
\begin{quote}
,,Bewirkt ein Strahlenb\"undel, da{\ss} ein von ihm getroffenes Molek\"ul die
Energiemenge~$h\,\nu$ in Form von Strahlung durch einen Elementarproze{\ss}
auf\-nimmt oder abgibt (Einstrahlung), so wird stets der
Impuls~$\frac{\displaystyle {h\,\nu}}{\displaystyle c}$ auf das Molek\"ul
\"ubertragen, und zwar bei der
Energieaufnahme in der Fortpflan\-zungs\-richtung des B\"undels, bei der
Energieabgabe in der entgegengesetzten Richtung. [...].''
\end{quote}
(A beam of light that
induces a molecule to absorb or deliver the energy~$h\,\nu$
as radiation by an elementary process (irradiation) will always transfer the
momentum~$\frac{\displaystyle {h\,\nu}}{\displaystyle c}$ to the molecule,
directed in the propagation direction of the beam for energy absorption, and
in the opposite direction for energy emission.)
\begin{quote}
,,Aber im allgemeinen begn\"ugt man sich mit der Betrachtung des
E\,n\,e\,r\,g\,i\,e-Austausches, ohne den I\,m\,p\,u\,l\,s-Austausch
zu ber\"ucksichtigen.''
\end{quote}
(However, in general one is satisfied with
the consideration of the e\,n\,e\,r\,g\,y exchange, without taking
the m\,o\,m\,e\,n\,t\,u\,m exchange into account.)

Let us first assume an atom~A with mass~$m$ in the ground state located
at the gravitational potential~$U_0 = 0$ and, therefore, with an energy
of~$E_0 = m\,c^2_0$. With an energy difference~$\Del E_0$ from the ground
state to the excited atom~A$^*$, the mass in this state
is~\cite{Ein05,Lau20,Lau11}:
%
%% Eq. 7
\begin{equation}
m + \Del m = \frac{1}{c^2_0}\,(E_0 + \Del E_0)~.
\label{Verwandlung}
\end{equation}
The masses $M$, $m$, and $\Del m$ constituting the total system considered
here are assumed to comply with the inequality $M \gg m \gg \Del m$, so that
higher orders can be neglected in some of the equations.
The ``rest energy'' with respect to the centre of gravity of $M$ and $m$
of the ground state at~$U$ will then be~\cite{Oku00}:
%
%% Eq. 8
\begin{equation}
E = E_0 + U\,m ~.
\label{Grund}
\end{equation}
The definition of the rest energy in this context calls for some further
explanations. If a particle with mass~$m$ is lowered from $U_0 = 0$ to~$U$,
the potential energy will be converted, for instance, into kinetic energy
of the particle, $E_{\rm kin} = - U\,m$. The total energy of the particle
at~$U$ will thus be $E_0 + E_{\rm kin}$. Provided the kinetic energy is
subsequently absorbed as thermal energy at~$U$, the remaining energy~$E_0$
of the particle---at rest with respect to the centre of gravity---is
obviously different from the rest energy in Eq.~(\ref{Grund}). The energy~$E_0$
will, however, not be available for any photon emission at~$U$, because
a lifting of the mass~$m$ to~$U_0$ would require the potential energy~$U\,m$,
whereas a photon would not change its energy during the transit from $U$ to
$U_0$, and could then be converted to mass there.
This accounts for the difference between~$E_0$ and the rest energy.

As will be
shown later, see, e.g., Eq.~(\ref{Paar}), momentum considerations also lead to
the requirement that only the \emph{rest energy} of Eq.~(\ref{Grund})
can be emitted as photon.

We now consider the rest energy~$E^*$ of the excited atom~A$^*$
at~$U$ and find
%
%% Eq. 9
\begin{eqnarray}
E^* = E_0 + \Del E_0 + U\,m + U\,\Del m ~,
\label{Energy}
\end{eqnarray}
where the remarks above apply as well.
%Again constraining forces are required and have transferred the
%energy~$- U\,(m + \Del m)$ to the total system.
In view of these energy equations, the transition of A$^*$ to the ground
state at~$U$ can provide an energy of
%
%% Eq. 10
\begin{equation}
\Del E =  E^* - E = \Del E_0 + U\,\Del m ~,
\label{difference}
\end{equation}
which is in principle available for the photon emission.
%(in the rest system
%of the centre of gravity of~$M$ and $m$, which basically the rest system
%of~$M$ with our assumption $M \gg m$).
Whether the emitted photon has the expected energy and frequency, can be
determined by observations; and the gravitational redshift measurements
mentioned in Sect.~\ref{s.cause} confirm indeed the right energy
%
%% Eq. 11
\begin{equation}
%E_{\nu'} =
\Del E = h\,\nu' ~,
\label{Frequenz}
\end{equation}
where $\nu'$ is measured with respect to the world time.

Nevertheless, the question remains \emph{how} the atom can sense the
potential~$U$ at the emission site and react accordingly. We will argue
that---in line with Einstein's remarks quoted---the momentum exchange
must be taken into account, in addition to the interaction of the radiation
energy with the potential energy of the emitting system. In preparation
for this task, we list some relevant relations.

The momentum of a photon emitted at~$U_0$ with frequency~$\nu_0$ is
%
%% Eq. 12
\begin{equation}
p_0 = \frac{h\,\nu_0}{c_0} = \frac{\Del E_0}{c_0}  ~,
\label{momentum_0}
\end{equation}
where $\Del E_0 = h\,\nu_0$ is its energy \cite{Ein17}.
At $U < 0$, the energy of the photon can be written as
%
%% Eq. 13
\begin{eqnarray}
% E_\nu = h\,\nu = p\,c
\Del E_0 = p\,c
% \frac{E_\nu}{c} =
% h\,\nu
\label{energy}
\end{eqnarray}
with a speed of light \cite{Oku00}
%
%% Eq. 14
\begin{equation}
c \approx c_0\,\left(1 + \frac{2\,U}{c^2_0}\right) ~.
\label{speed}
\end{equation}

This speed is in agreement with an evaluation by Schiff for radial propagation
in a central gravitational field \cite{Sch60}. A decrease of the speed of
light near the Sun of this amount is not only supported by the predicted and
subsequently observed Shapiro delay \cite{Sha71,Kraetal}, but also
indirectly by the deflection of light \cite{Ein16,Dys20}.

The problem can then be illustrated by different scenarios for the emission
process:
\begin{enumerate}
\item[(a)]

Under the assumption that the atom can somehow locally sense the gravitational
potential~$U$, but not the speed~$c$, the energy given by
Eq.~(\ref{difference}) would lead to a momentum
%
%% Eq. 15
\begin{equation}
p = \frac{\Del E}{c_0} = \frac{\Del E_0 + U\,\Del m}{c_0}
\label{p_a}
\end{equation}
of the photon after the emission. We could then estimate its
energy by applying Eqs.~(\ref{energy}) and (\ref{speed})
%
%% Eq. 16
\begin{equation}
p\,c \approx \frac{\Del E_0 + U\,\Del m}{c_0}\,c_0\,\left(1 + \frac{2\,U}
{c^2_0}\right) \approx \Del E_0 + 3\,U\,\Del m ~,
\label{E_a}
\end{equation}
with $\Del E_0/c^2_0 = \Del m $ according to Eq.~(\ref{Verwandlung}),
and neglecting higher orders of $U/c^2_0$.
The energy thus obtained is in conflict with Eq.~(\ref{difference}).

\item[(b)]
If the atom can, however, sense the local speed of light~$c$, but not
the potential~$U$, the photon emission energy will be~$\Del E_0$, which
is also in conflict with Eq.~(\ref{difference}).

\item[(c)]
If the atom can sense both the speed of light~$c$ and
the potential~$U$, it then has to reduce the photon
emission energy by a factor of~$(1 + U/c^2_0)$ and, at the same time, increase
the photon momentum by a factor of~$(1 - U/c^2_0)$. Although this scenario
is formally correct, it involves very unlikely processes.

\item[(d)]
If Einstein's assumption that only intra-atomic processes are of importance is
valid, this is equivalent to the statement that the atom can sense neither~$U$
nor~$c$. The internal transition of A$^*$ to the ground state of atom~A then
proceeds in the same way at~$U_0$ and $U$; in both cases, accompanied by an
energy release of~$\Del E_0$ and a momentum of~$\Del E_0/c_0$. The adjustment
of the energy and momentum transfers to the rest system of the centre of
gravity will be achieved during the actual photon emission at the speed~$c$,
as will be detailed below.

\end{enumerate}

The intra-atomic processes are indicated in rows~2 to 4 of Table~1.
Starting from an excited atom~A$^*$ at $U$, the transition energy and
momentum are given according to Eqs.~(\ref{Verwandlung}) and
(\ref{momentum_0}).
We argue that only the propagation speed~$c$ of photons in the environment of
the emission location provides the necessary information for the energy and
momentum adjustments in line with the corresponding conservation laws.

The sequence of events will be modelled according
to an explanation of the Doppler effect based on energy and
momentum conservations by \citet{Fer32}, which has some resemblance
to the Compton effect~\cite{Com23}. Fermi discussed the interaction
of the liberated energy during an atomic transition with the \emph{kinetic
energy} of the emitter and its momentum in a non-relativistic
approximation.

In our case, the interactions of the \emph{potential energy} and momentum
during the emission of a photon can be formulated by the introduction of an
arbitrary differential momentum vector~$\vec{x}$ parallel to $\vec{p_0}$,
which has to be determined by solving the momentum and energy equations of the
atom-photon system in rows 6 and 7 of Table~1. Row~6 is clearly
consistent with momentum conservation and row~7 leads to
%
%% Eq. 17
\begin{equation}
\Del E_0 - ||\vec{x}||\,c_0 = ||\vec{p_0} - \vec{x}||\,c_0 = p\,c =
||\vec{p_0} + \vec{x}||\,c
\label{Energy_2}
\end{equation}
for the energy relationship. The kinetic energy~$E_{\rm kin}$, the recoil
energy, can be neglected, because it is already very small with our
assumption~$m >> \Del m$, but has been further
reduced in the Pound--Rebka-experiment \cite{PouReb} with the help of the
M\"o{\ss}bauer effect \cite{Moe58}.
From Eq.~(\ref{Energy_2}), it follows with Eq.~(\ref{speed})
%
%% Eq. 18
\begin{equation}
\frac{p_0 - x}{p_0 + x} = \frac{c}{c_0} \approx 1 + \frac{2\,U}{c^2_0} ~,
\label{Speed}
\end{equation}
where $p_0 = ||\vec{p_0}||$ and  $x = ||\vec{x}||$. The evaluation yields in our
approximation
%
%% Eq. 19
\begin{equation}
x \approx - p_0\,\frac{U}{c^2_0} ~.
\label{Correction}
\end{equation}
Hence, we get for the momentum of the photon
%
%% Eq. 20
\begin{equation}
p \approx p_0\,\left(1 - \frac{U}{c^2_0}\right) ~.
\label{momentum}
\end{equation}
The result is that $p$ will be larger than $p_0$. This can be
understood by considering that the energy transfer of $||\vec{x}||\,c_0$ in
Eq.~(\ref{Energy_2}) back to the atom in the gravitational field of the
mass~$M$ must be accompanied by a momentum transfer of $p_0\,U/c^2_0$
and a corresponding reaction on the photon in line with Eq.~(\ref{momentum}).
Note that the energy transfer~$x\,c_0 = - p_0\,U/c_0 = - U\,\Del m$
is of the same amount as the difference of potential energy gains by
lowering $m + \Del m$ and $m$ in the field.  Taking the remarks related to
Eqs.~(\ref{Grund}) and (\ref{Energy}) into account, the energy levels before
the emission of the photon are
$\Del E_0 = \Del m\,c^2_ 0$ at $U_0 = 0$ and
%
%% Eq. 21
\begin{equation}
\Del E = \Del m\,c^2_ 0 - U\,\Del m
\label{Energy_level}
\end{equation}
at~$U$, where~$- U\,\Del m$ is the potential energy at~$U_0$ relative to~$U$
converted, for instance, into kinetic energy of the atom. Assuming it is
brought to a halt by constraining forces, an
energy~$\Del E' = \Del E_0 = \Del m\,c^2_ 0$ remains. As we have seen, it
cannot directly be converted into energy, because of momentum considerations,
but
%
%% Eq. 22
\begin{equation}
h\,\nu = \Del E_0\,\left(1 + \frac{U}{c^2_0}\right) =
\Del m\,c^2_ 0 + U\,\Del m
\label{Emitted}
\end{equation}
can be emitted and can propagate to~$U_0$. The conversion of~$\Del m$ into
energy entails a loss of the potential energy gain of~$-U\,\Del m$ mentioned
above. It will be replenished by the energy transfer~$x\,c_0$.
The energy budget after the photon emission then is
$\Del m\,c^2_ 0 + U\,\Del m$ at~$U_0$ plus $- 2\,U\,\Del m$ at~$U$ giving a
total of $ \Del m\,c^2_ 0 - U\,\Del m$ in agreement with
Eq.~(\ref{Energy_level}).
%increases the
%potential energy of the emitter by~$- U\,\Del m$ as it should be in order to
The gravitational redshift in Eq.~(\ref{Emitted}) is consistent with
Eq.~(\ref{shift}) and observations.

\subsection{The Compton frequency controversy of Wolf \emph{et al.} and
M\"uller \emph{et al.}}%%%%%%%%%%%%%%%%%%%%%%%%%%%%%%%%%%%%%%%%%%%%%%%%%%%%%%
\label{Compton}
%% Subsect. III B

In a formal way, we can also compare~$E^*$ of Eq.~(\ref{Energy}) with
%
%% Eq. 23
\begin{eqnarray}
E_1 = E_0 + U_1\,m ~,
\label{Delta_pot}
\end{eqnarray}
the rest energy of the ground state at a different
potential~$U_1 = U + \delta U$ at a position close to that of the
potential~$U$.
If $U_1$ is chosen such that
%
%% Eq. 24
\begin{equation}
U_1\,m = U\,(m + \Del m) ~,
\label{Definition}
\end{equation}
subtraction of Eq.~(\ref{Delta_pot}) from Eq.~(\ref{Energy}) gives
%
%% Eq. 25
\begin{eqnarray}
E^* - E_1
 = \Del E_0 ~,
\label{Delta_E_0}
\end{eqnarray}
which suggests that the energy~$\Del E_0$ would be available assuming
a more or less instantaneous shift of the atom from $U$ to $U_1$. This is,
however, not possible. The selection of $U_1$ in Eq.~(\ref{Definition}),
nevertheless, leads to the interesting relation
%
%% Eq. 26
\begin{eqnarray}
U\,\Del m = m\,\delta U  ~,
\label{Beziehung}
\end{eqnarray}
which shows that the energy difference will be determined by the
gravitational potential, if a mass variation~$\Del m$ is involved. On the
other hand, the potential difference~$\delta U$ is of importance, if the
emitter with mass~$m$ changes its position.
In this sense, both statements \cite{Wol10,Mue10b}
cited above contain some truth. It would, however, be required to
formulate the corresponding premises in great detail.

\subsection{Pair annihilation}%%%%%%%%%%%%%%%%%%%%%%%%%%%%%%%%%%%%%%%%%%%%%%%
\label{solution_pairs}
%% Subsect. III C

We first formulate the rest energy
of both particles involved---here an electron and a
positron---at the gravitational potential~$U$ as
%
%% Eq. 27
\begin{equation}
2\,E^\pm = 2\,E^\pm_0 + 2\,U\,m_{\rm e}
\label{Paar}
\end{equation}
with rest energies of $E^\pm_0 = m_{\rm e}\,c^2_0$ at~$U_0 = 0$.
We will neglect any transitions from its excited states and
assume a final state that eventually disintegrates into
two $\gamma$-ray photons of equal energy~$E$, but in opposite
directions \cite{Smi02}.
In a formal way, in analogy to Sect.~\ref{solution_lines},
each photon can only get half the energy given by Eq.~(\ref{Paar}) in the
rest system of the centre of gravity.

As for the photon emission of an atomic particle, the question arises which
parameter controls this emission energy. The answer again is that the
speed of light~$c$ at~$U$ is the decisive factor.
In Table~2 are summarized the momentum and energy terms---written
under the assumption that the initial annihilation is not dependent on the
gravitational potential~$U$, but the emission process of the photons is
affected by the speed of light in accordance with the results in
Sect.~\ref{solution_lines}. The momentum conservation follows from the
symmetry of the emissions. The energy equations for each of the photons in
line with energy conservation can be written as
%
%% Eq. 28
\begin{equation}
E^\pm_0 - X\,c_0 = (P_0 - X)\,c_0 = (P_0 + X)\,c = h\,\nu ~,
\label{Energy_3}
\end{equation}
where $P_0 = ||\pm \vec{P_0}||$, $X = ||\pm \vec{X}||$, and $\pm \vec{X}$ are
arbitrary differential momentum vectors parallel to $\pm \vec{P_0}$,
which have to be determined by solving Eq.~(\ref{Energy_3})
related to row~8 of Table~2.
With Eq.~(\ref{speed}) it follows
%
%% Eq. 29
\begin{equation}
\frac{P_0 - X}{P_0 + X} = \frac{c}{c_0} \approx 1 + \frac{2\,U}{c^2_0}
\label{Speed_result}
\end{equation}
and
%
%% Eq. 30
\begin{equation}
X = - P_0\,\frac{U}{c^2_0} ~.
\label{correction}
\end{equation}
Notice, in this case, that the energy ~$2\,X\,c_0 =
- 2\,P_0\,U/c_0$ corresponds to the potential
energy~$- 2\,U\,m_{\rm e}$ of the electron and positron at $U_0$ with
respect to $U$.

The same arguments as those for spectral lines in Sect.~\ref{solution_lines}
then result in a
relative gravitational redshift consistent with Eq.~(\ref{shift}).

\section{Conclusion}%%%%%%%%%%%%%%%%%%%%%%%%%%%%%%%%%%%%%%%%%%%%%%%%%%%%%%%%%
\label{concl}
%% Sect. IV

In summary, it can be concluded that the internal processes of an atom or ion
during transitions between different energy states
will not be significantly influenced by a moderate gravitational field, but
the conversion of the liberated energy into a photon will be affected by the
local gravitational potential via the speed of light
and gives the observed redshift. Matter-antimatter
pair annihilation leads to the same relative redshift, albeit with a slightly
different interaction process in the near-field radiation region.

%%%%%%%%%%%%%%%%%%%%%%%%%%%%%%%%%%%%%%%%%%%%%%%%%%%%%%%%%%%%%%%%%%%%%%%%%%%%%
%%%%%%%%%%%%%%%%%%%%%%%%%%%%%%%%%%%%%%%%%%%%%%%%%%%%%%%%%%%%%%%%%%%%%%%%%%%%%
%%%%%%%%%%%%%%%%%%%%%%%%%%%%%%%%%%%%%%%%%%%%%%%%%%%%%%%%%%%%%%%%%%%%%%%%%%%%%

%\begin{acknowledgements}
Acknowledgements:
We thank two anonymous reviewers for their constructive comments on the
manuscript.
This research has made extensive use of
the Astrophysics Data System (ADS).
%\end{acknowledgements}

%\begin{acknowledgements}

%\end{acknowledgements}

%% The Appendices part is started with the command \appendix;
%% appendix sections are then done as normal sections
%% \appendix

%% \section{}
%% \label{}

%% References
%%
%% Following citation commands can be used in the body text:
%% Usage of \cite is as follows:
%%   \cite{key}         ==>>  [#]
%%   \cite[chap. 2]{key} ==>> [#, chap. 2]
%%

%% References with BibTeX database:

%\bibliographystyle{elsarticle-num}
%\bibliography{<your-bib-database>}

%% Authors are advised to use a BibTeX database file for their reference list.
%% The provided style file elsarticle-num.bst formats references in the required Procedia style

%% For references without a BibTeX database:

% \begin{thebibliography}{00}

%% \bibitem must have the following form:
%%   \bibitem{key}...
%%

% \bibitem{}

% \end{thebibliography}

%%
%% End of file `ecrc-template.tex'.

%
%% Table 1
\begin{table}[b]
%\begin{footnotesize}
\begin{tabular}[t]{|c|c|cc|c|} \hline
1& & Transition of & Atom~A$^*$ at $U$ &  \\ \hline  \hline
2& Energy & $\Del m\,c^2_0 = \Del E_0~~~=$ &
$||\vec{p_0}||\,c_0$ & see Eqs.~(\ref{Verwandlung}) \\
3& Momentum: & $- \vec{p_0}$ &  $\vec{p_0}$ & and~(\ref{momentum_0}) \\
4& direction &$\longleftarrow$ & $\longrightarrow$ &  \\ \hline
5&$\longleftarrow\,\leftarrow$ &  & $\leftarrow$ &
$\Longrightarrow\,\Rightarrow$ \\
6&$- \vec{p_0} - \vec{x}$& &$- \vec{x}$&$\vec{p_0} + \vec{x}$ \\
7&$-U\,\Del m$ &$E_{\rm kin} \ll \Del E_0$ & $||\vec{p_0} - \vec{x}||\,c_0$ &
$||\vec{p_0} + \vec{x}||\,c$ \\  \hline \hline
8&Atom~A & Interaction & region & Photon \\ \hline
\end{tabular}

\vspace{0.3cm}

Table~1: \quad Transition of an excited atom to the ground state at a
gravitational potential~$U$. In rows~2 to 4 of the central column---called
``Interaction region''---the left-hand side is related to the atom and the
right-hand side refers to the near-field radiation during the emission process,
which, according to Einstein's early assumption quoted above, is controlled by
the atom alone and therefore does not dependent on~$U$. In rows~5
to 7, the photon emission and the reaction onto the emitter are indicated in
line with momentum and energy conservation, cf., Eqs.~(\ref{Correction}) and
(\ref{Energy_2}). The M\"o{\ss}bauer effect can be employed to increase the
mass of the emitter and allow us to neglect~$E_{\rm kin}$. The momentum
vectors are drawn by solid arrows, whereas the propagating photon is
characterized by open momentum arrows.
%\end{footnotesize}
\label{tab_1}
\end{table}
%
%
%% Table 2
\begin{table}[t]
%\begin{footnotesize}
\begin{tabular}{|c|c|cc|c|} \hline
1&Electron & $E^{-}_0 = m_{\rm e}\,c^2_0$~~&$E^{+}_0 = m_{\rm e}\,c^2_0$ &
Positron \\ \hline \hline
2& Energy & $||- \vec{P_0}||\,c_0$ &  $||+ \vec{P_0}||\,c_0$ &
cf., Eq.~(\ref{momentum_0}) \\
3& Momentum & $- \vec{P_0}$ &  $+ \vec{P_0}$ & \\
\hline
4& & $\longleftarrow$ & $\longrightarrow$ &\\
5& & $\rightarrow$ & $\leftarrow$ & \\
6& $\Leftarrow$ $\Longleftarrow$ & & & $\Longrightarrow$ $\Rightarrow$ \\
\hline
7&$- \vec{P_0} - \vec{X}$&$+ \vec{X}$&$- \vec{X}$&$+ \vec{P_0} + \vec{X}$ \\
8&$~||-\vec{P_0} - \vec{X}||\,c~$ & $~||-\vec{P_0} + \vec{X}||\,c_0~$ &
$~||+\vec{P_0} - \vec{X}||\,c_0~$ &
$~||+ \vec{P_0} + \vec{X}||\,c~$ \\  \hline  \hline
9&Photon~1 & Interaction &region~~~~& Photon~2 \\ \hline
\end{tabular}

\vspace{0.3cm}

Table~2: \quad Pair annihilation of an electron and a positron at a
gravitational potential~$U$. The table is structured similar to Table~1, but
the near-field interaction region now concerns the momentum and energy
relationships during the emissions of the photons~1
and 2. In rows 7 and 8, the momentum and energy relationships are indicated,
cf., Eqs.~(\ref{correction}) and (\ref{Energy_3}).
As in Table~1, the momentum vectors are
drawn by solid arrows, whereas the propagating photons are characterized by
open momentum arrows.
%\end{footnotesize}
\label{tab_2}
\end{table}
\end{document}